\documentclass[twocolumn,showpacs,preprintnumbers,amsmath,amssymb]{revtex4}
\usepackage{graphicx}% Include figure files
\usepackage{dcolumn}% Align table columns on decimal point
\usepackage{bm}% bold math

\begin{document}

\preprint{}

\title{Quantized photonic spin Hall effect in graphene}% Force line breaks with \\
\author{Liang Cai}
%\altaffiliation[ ]{}%Lines break automatically or can be forced with \\
\author{Mengxia Liu}
\author{Shizhen Chen}
\author{Yachao Liu}
\author{Weixing Shu}
\author{Hailu Luo}\email{hailuluo@hnu.edu.cn}
\author{Shuangchun Wen}
%%\email{Second.Author@institution.edu}
\affiliation{Laboratory for spin photonics, School of Physics and Electronics, Hunan University, Changsha 410082,China}
\date{\today}% It is always \today, today,

\begin{abstract}
We examine the photonic spin Hall effect (SHE) in a graphene-substrate system with the presence of external magnetic field.
In the quantum Hall regime, we demonstrate that the in-plane and transverse spin-dependent splittings in photonic SHE exhibit different quantized behaviors.
The quantized SHE can be described as a consequence
of a quantized geometric phase (Berry phase), which
corresponds to the quantized spin-orbit interaction.
Furthermore, an experimental scheme based on quantum weak value amplification is proposed to detect
the quantized SHE in terahertz frequency regime.
By incorporating the quantum weak measurement techniques, the quantized photonic SHE
holds great promise for detecting quantized Hall conductivity and Berry phase.
These results may bridge the gap between the electronic SHE and photonic SHE in graphene.
\end{abstract}


%Use showkeys class option if keyword
\maketitle

\section{Introduction}\label{SecI}
The spin Hall effect (SHE) is a transport phenomenon in
which an electric field applied to spin particles leads to a spin-dependent
displacement perpendicular to the direction of electric field ~\cite{Murakami2003,Sinova2004,Wunderlich2005}.
The SHE in electronic system offers an effective way to manipulate
the spin particles and therefore opens a promising way to potential applications in
semiconductor spintronic devices~\cite{Wolf2001}. Graphene has emerged as an attractive
material for spintronics due to long spin diffusion
lengths~\cite{Tombros2007}, high-efficiency spin injection~\cite{Han2010}, and
tunable spin transport~\cite{Dushenko2016}. Recently, a quantized SHE for electron is shown
to occur in graphene due to spin-orbit interaction plus the peculiarity of
graphene electronic structure~\cite{Kane2005I}. The quantized electronic SHE has also been predicted
to occur in graphene without spin-orbit interaction but with an external
magnetic field~\cite{Abanin2006}. In general, the quantum SHE is different from quantized SHE, due to
the fact that the quantum spin Hall phase is not generally characterized by a
quantized spin Hall conductivity~\cite{Kane2005II}.

The photonic SHE can be regarded as a direct
optical analogy of SHE in electronic system where the spin electrons
and electric potential are replaced by spin photons and refractive
index gradient, respectively~\cite{Onoda2004,Bliokh2006,Hosten2008}.
The photonic SHE is
generally considered as a result of an effective spin-orbit
interaction, which describes the mutual influence of the spin
(polarization) and trajectory of the light beam.
In past few years the photonic SHE has been intensively investigated in
different physical systems, such as high-energy physics~\cite{Gosselin2007,Dartora2011},
plasmonics~\cite{Gorodetski2008,Yin2013,Kapitanova2014,Xiao2015}, optical physics~\cite{Bliokh2008,Haefner2009,Aiello2009,Herrera2010,Qin2011,Korger2014,Ling2015}, and semiconductor
physics~\cite{Menard2010}. Recently, the photonic SHE has been
examined in graphene~\cite{Zhou2012} and topological insulators~\cite{Zhou2013}.
It has been demonstrated that the photonic SHE is a useful metrological tool to characterize the structure
parameters of two-dimensional atomic crystals. More recently, the quantized Imbert-Fedorov effect and Goos-H\"{a}nchen effect
have been theoretically predicted in quantum Hall regime of graphene-substrate system~\cite{Kamp2016}.
Interestingly, the relative photonic SHE only depends on the optical properties of substrate,
and therefore is not quantized.
In general, the Imbert-Fedorov effect and photonic SHE are two different effects,
although they arise from the same physical root: spin-orbit interaction of light~\cite{Bliokh2013,Bliokh2015I}.

In this paper, we examine the photonic SHE in quantum Hall regime of graphene.
Here electron-phonon interaction and impurity scattering are neglected
since they can destroy the quantization of the Hall conductivity~\cite{Li2013I,Li2014I}.
First, we develop a general model to describe the photonic SHE in reflection at the graphene-substrate interface.
Based on this model, both in-plane and transverse spin-dependent splitting in photonic SHE are obtained. Next, we attempt to reveal the quantized photonic SHE in quantum Hall regime of graphene.
We demonstrate that the in-plane and the transverse splittings in
photonic SHE exhibit different quantized behaviors.
The quantized SHE can be described as a consequence
of a quantized geometric phase (Berry phase), which
corresponds to the quantized spin-orbit interaction.
Finally, we propose a scheme known as quantum weak measurements to detect the
quantized photonic SHE in terahertz frequency regime.

\section{General theoretical model}\label{SecII}
\begin{figure}
\centerline{\includegraphics[width=8cm]{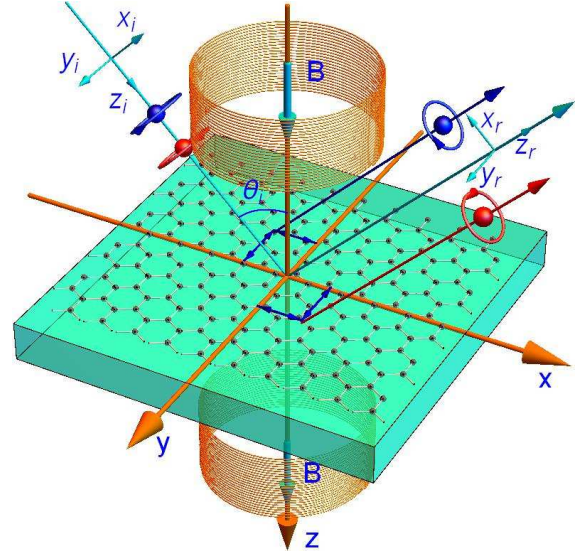}}
\caption{\label{Fig1}(color online) Schematic illustrating the
wave reflection at a graphene-substrate
interface in Cartesian coordinate system. The graphene
sheet is placed on the top of a homogeneous and isotropic substrate. An imposed static magnetic field $\mathbf{B}$ is applied perpendicular
to the interface of the graphene-substrate system. On the reflecting surface,
the photonic SHE occurs which manifests itself as in-plane and transverse splittings of two spin components.}
\end{figure}

We first establish a general model to describe the photonic SHE in the graphene-substrate system.
Let us consider a Gaussian wavepacket with monochromatic frequency $\omega$ impinging from air to graphene-substrate
system as shown in Fig.~\ref{Fig1}. The $z$ axis of the laboratory Cartesian frame ($x,y,z$) is normal to the graphene-substrate
system. A graphene sheet is placed on the top of substrate,
and a static magnetic field $\mathbf{B}$ is applied along the $z$ axis.
In addition, we use the coordinate frames ($x_i,y_i,z_i$) and
($x_r,y_r,z_r$) to denote central
wave vector of incidence and reflection, respectively.
For horizontal polarization $|{H}({k}_{i,r})\rangle$ and vertical polarization $|{V}({k}_{i,r})\rangle$
the corresponding individual wave-vector components can be expressed by $|{P}({k}_{i}\rangle$
 and $|{S}({k}_{i})\rangle$~\cite{Onoda2004,Bliokh2006,Hosten2008}:
\begin{equation}
 |{H}(k_{i,r})\rangle=|P({k}_{i,r})\rangle-\frac{k_{iy}}{k_{i,r}}\cot\theta_{i,r}|\mathbf{S}({k}_{i,r})\rangle\label{HKIR},
 \end{equation}
\begin{equation}
 |{V}({k}_{i,r})\rangle=|{S}(\mathbf{k}_{i,r})\rangle+\frac{k_{iy}}{k_{i,r}}\cot\theta_{i,r}|{P}({k}_{i,r})\rangle\label{VKIR},
\end{equation}
 where $\theta_i$ and $\theta_r$ denote the incident and reflected angles,
 $k_i$ and $k_r$ are the incident and reflected wavevectors, respectively.
 After reflection, $|P({k}_{i})\rangle $ and $|{S}({k}_{i})\rangle $ evolve as $[|{P}({k}_r)\rangle~|{S}({k}_r)\rangle]^T={m}_{R}[|{p}({k}_i)\rangle~|{s}({k}_i)\rangle]^T$, where
\begin{eqnarray}{m}_{R}=
\left[
\begin{array}{cc}
r_{pp} &r_{ps}\\
r_{sp} &r_{ss}
\end{array}\right]\label{MatrixRI}.
\end{eqnarray}
Based on the boundary conditions, the Fresnel reflection coefficients of the graphene-substrate system in the presence of an imposed magnetic field can be obtained as~\cite{Tse2011,Kamp2015}
  \begin{equation}
  r_{pp}=\frac{\alpha^T_+\alpha_-^L+\beta}{\alpha_+^T\alpha_+^L+\beta}\label{RPP},
  \end{equation}
  \begin{equation}
  r_{ss}=-\frac{\alpha^T_-\alpha_+^L+\beta}{\alpha^T_+\alpha_+^L+\beta}\label{RSS},
  \end{equation}
   \begin{equation}
   r_{ps}=r_{sp}=-2\sqrt{\frac{\mu_0}{\varepsilon_0}}\frac{k_{iz}k_{tz}\sigma_H}{\alpha^T_+\alpha_+^L+\beta}\label{RPS}.
   \end{equation}
  Here, $\alpha^L_\pm=(k_{iz}\varepsilon\pm k_{tz}\varepsilon_0+k_{iz}k_{tz}\sigma_L/\omega)/\varepsilon_0$,
  $\alpha^T_\pm=k_{tz}\pm k_{iz}+\omega\mu_0\sigma_T$,
  $\beta=\mu_0k_{iz}k_{tz}\sigma^2_H/\varepsilon_0$,
$k_{iz}=k_i\cos\theta_i$, and $k_{tz}=k_t \cos\theta_t$;
  $\theta_t$ is the refraction angle; $\varepsilon_0$ , $\mu_0$  are
  permittivity and permeability in vacuum, respectively; $\varepsilon$ is the permittivity of substrate;
  $\sigma_L$, $\sigma_T$ and $\sigma_H$
  denote the longitudinal, transverse, and Hall conductivity, respectively.

  When the applied magnetic field is strong enough,
  the Hall conductivity  in quantum Hall regime of the graphene is quantized in multiples of the fine structure constant and is given by~\cite{Kamp2015}
  \begin{equation}
  \sigma_H=2(2n_c+1)\mathrm{Sgn}[B]\frac{e^2}{2\pi\hbar}\label{HallC}.
  \end{equation}
  Here, $n_c=\mathrm{Int}[\mu^2_F/2\hbar e|B|v^2_F]$
  is the number of occupied Landau levels, $\mu_F$ and $v_F$ are the  Fermi energy and the Fermi velocity, respectively.
However, Eqs~(\ref{HallC}) no longer holds for intermediate or weak magnetic fields~\cite{Kamp2016}.
Note that in the visible region of the spectrum an important model of Fresnel coefficients in graphene has been developed by fixing both the surface susceptibility and the
surface conductivity~\cite{Merano2016I,Merano2016II}.

From Eqs.~(\ref{HKIR})-(\ref{MatrixRI}),
the reflected polarization states for different angular spectrum components can be written as
$[|{{H}}({k}_r)\rangle~|{{V}}({k}_r)\rangle]^T={M}_{R}[|{H}({k}_i)\rangle~|{V}({k}_i)\rangle]^T$~\cite{Luo2011I}.
Here, ${M}_{R}$ can be expressed as
\begin{eqnarray}
\left[
\begin{array}{cc}
r_{pp} &r_{ps}+\frac{k_{ry}\cot\theta_{i}(r_{pp}+r_{ss})}{k_{0}} \\
r_{sp}-\frac{k_{ry}\cot\theta_{i}(r_{pp}+r_{ss})}{k_{0}} &
r_{ss}
\end{array}\right]\label{MatrixRII}.
\end{eqnarray}
In above equation, we have introduced the boundary condition
$k_{rx}=-k_{ix}$ and $k_{ry}= k_{iy}$. By making use of Taylor
series expansion based on the arbitrary angular spectrum component,
$r_{ab}$ can be expanded as a polynomial of $k_{ix}$:
\begin{eqnarray}
r_{ab}(k_{ix})&=&r_{ab}(k_{ix}=0)+k_{ix}\left[\frac{\partial
r_{p,s}(k_{ix})}{\partial
k_{ix}}\right]_{k_{ix}=0}\nonumber\\
&&+\sum_{j=2}^{N}\frac{k_{ix}^j}{j!}\left[\frac{\partial^j
r_{ab}(k_{ix})}{\partial k_{ix}^j}\right]_{k_{ix}=0}\label{Talorkx},
\end{eqnarray}
where $r_{ab}$ denotes the Fresnel reflection coefficients with $a$ and $b$
standing for either $s$ or $p$ polarization. The polarizations associated
with each angular spectrum components experience different rotations
in order to satisfy the boundary condition after reflection:
\begin{eqnarray}
 |{H}({k}_{i})\rangle&\rightarrow& r_{pp}|{H}({k}_{r})\rangle-\frac{k_{ry}\cot\theta_{i}(r_{pp}+r_{ss})}{k_{0}}|{V}({k}_{r})\rangle\nonumber\\
 &&+ r_{sp}|{V}({k}_{r})\rangle-\frac{k_{rx}}{k_0}\frac{\partial r_{sp}}{\partial \theta_i}|{V}({k}_{r})\rangle\label{HKI},
\end{eqnarray}
\begin{eqnarray}
 |{V}({k}_{i})\rangle&\rightarrow& r_{ss}|{V}({k}_{r})\rangle+\frac{k_{ry}\cot\theta_{i}(r_{pp}+r_{ss})}{k_{0}}|{H}({k}_{r})\rangle\nonumber\\
 &&+r_{ps}|{H}({k}_{r})\rangle-\frac{k_{rx}}{k_0}\frac{\partial r_{ps}}{\partial \theta_i}|{H}({k}_{r})\rangle
 \label{VKI}.
\end{eqnarray}
To accurately describe the photonic SHE, the Fresnel
reflection coefficients are confined to the first order in Taylor series expansion.

\section{Quantized photonic spin Hall effect}\label{SecIII}
It is well known that the photonic SHE manifests itself as
spin-dependent splitting which appears in both position and momentum spaces.
To reveal the photonic SHE,
we now determine the in-plane and transverse shifts of the wavepacket.
In the spin basis set, the polarization state of $|{H}\rangle$ or $|{V}\rangle$ can be decomposed into two orthogonal spin components
\begin{equation}
|H\rangle=\frac{1}{\sqrt{2}}(|{+}\rangle+|{-}\rangle),\label{SBH}
\end{equation}
\begin{equation}
|V\rangle=\frac{1}{\sqrt{2}}i(|{-}\rangle-|{+}\rangle),\label{SBV}
\end{equation}
where $|{+}\rangle$ and $ |{-}\rangle$ represent the left- and right-circular polarization components, respectively.

From Eqs.~(\ref{HKI}) and (\ref{VKI}), the reflected polarization states
 $|{\psi}^{H}_{r}\rangle$ and $|{\psi}^{V}_{r}\rangle$ in the momentum space can be obtained as
 \begin{eqnarray}
  |{\psi}^{H}_{r}\rangle&=&\frac{r_{pp}-ir_{sp}}{\sqrt{2}}(1+ik_{rx}\delta_{rx+}^H+ik_{ry}\delta_{ry+}^H) |+\rangle\nonumber\\
  &&+\frac{r_{pp}+ir_{sp}}{\sqrt{2}}(1-ik_{rx}\delta_{rx-}^H-ik_{ry}\delta_{ry-}^H)|-\rangle,\nonumber\\\label{WFH}
  \end{eqnarray}
  \begin{eqnarray}
    |{\psi}^{V}_{r}\rangle&=&\frac{r_{ps}-ir_{ss}}{\sqrt{2}}(1+ik_{rx}\delta_{rx+}^V+ik_{ry}\delta_{ry+}^V) |+\rangle\nonumber\\
  &&+\frac{r_{ps}+ir_{ss}}{\sqrt{2}}(1-ik_{rx}\delta_{rx-}^V-ik_{ry}\delta_{ry-}^V)|-\rangle.\nonumber\\\label{WFH}
  \end{eqnarray}
 Here, $\delta_{rx\pm}^H=(\partial r_{sp}/\partial \theta_i)/[(r_{pp}\mp ir_{sp})k_0]$, $\delta_{ry\pm}^H=[(r_{pp}+r_{ss})\cot\theta_i]/[(r_{pp}\mp ir_{sp})k_0]$,
 $\delta_{rx\pm}^V=(-\partial r_{ps}/\partial \theta_i)/[(r_{ss}\pm i r_{ps})k_0]$,
$\delta_{ry\pm}^V=[(r_{pp}+r_{ss})\cot\theta_i]/[(r_{ss}\pm ir_{ps})k_0]$, and $k_0=\omega/c$ is the wavevector in vacuum.

We assume that the wavefunction in momentum space can be specified by the following expression
\begin{equation}
|\Phi\rangle=\frac{w_{0}}{\sqrt{2\pi}}\exp\left[-\frac{w^{2}_{0}(k_{ix}^{2}+k_{iy}^{2})}{4}\right]\label{GaussianWF},
\end{equation}
where $w_{0}$ is the width of wavefunction. The total wavefunction is made up of the packet spatial extent and
the polarization description:
 \begin{eqnarray}
   |{\Phi}^{H}_{r}\rangle&=&\bigg[\frac{r_{pp}-ir_{sp}}{\sqrt{2}}(1+ik_{rx}\delta_{rx+}^H+ik_{ry}\delta_{ry+}^H) |+\rangle\nonumber\\
  &&+\frac{r_{pp}+ir_{sp}}{\sqrt{2}}(1-ik_{rx}\delta_{rx-}^H-ik_{ry}\delta_{ry-}^H)\nonumber\\&&\times|-\rangle\bigg]|\Phi\rangle,\label{WPHI}
  \end{eqnarray}
  \begin{eqnarray}
     |{\Phi}^{V}_{r}\rangle&=&\bigg[\frac{r_{ps}-ir_{ss}}{\sqrt{2}}(1+ik_{rx}\delta_{rx+}^V+ik_{ry}\delta_{ry+}^V) |+\rangle\nonumber\\
  &&+\frac{r_{ps}+ir_{ss}}{\sqrt{2}}(1-ik_{rx}\delta_{rx-}^V-ik_{ry}\delta_{ry-}^V)\nonumber\\&&\times|-\rangle\bigg]|\Phi\rangle.\label{WPVI}
  \end{eqnarray}

If the spin-orbit interaction at the interface
 reflection is weak, Eqs.~(\ref{WPHI}) and (\ref{WPVI}) can be written as
 \begin{eqnarray}
  |{\Phi}^{H}_{r}\rangle&=&\bigg[\frac{r_{pp}-ir_{sp}}{\sqrt{2}}\exp(+ik_{rx}\delta_{rx+}^H)\exp(+ik_{ry}\delta_{ry+}^H)] |+\rangle\nonumber\\
  &&+\frac{r_{pp}+ir_{sp}}{\sqrt{2}}\exp(-ik_{rx}\delta_{rx-}^H)\exp(-ik_{ry}\delta_{ry-}^H)\nonumber\\&&\times|-\rangle\bigg]|\Phi\rangle,\label{WPHII}
  \end{eqnarray}
  \begin{eqnarray}
    |{\Phi}^{V}_{r}\rangle&=&\bigg[\frac{r_{ps}-ir_{ss}}{\sqrt{2}}\exp(+ik_{rx}\delta_{rx+}^V)\exp(+ik_{ry}\delta_{ry+}^V) |+\rangle\nonumber\\
  &&+\frac{r_{ps}+ir_{ss}}{\sqrt{2}}\exp(-ik_{rx}\delta_{rx-}^V)\exp(-ik_{ry}\delta_{ry-}^V)\nonumber\\&&\times|-\rangle\bigg]|\Phi\rangle.\label{WPVII}
  \end{eqnarray}
Here, the approximations of $1+i\sigma k_{rx}\delta^{H,V}_{rx\pm}\approx\exp(i\sigma k_{rx}\delta^{H,V}_{rx\pm})$ and $1+i\sigma k_{ry}\delta^{H,V}_{ry\pm}\approx\exp(i\sigma k_{ry}\delta^{H,V}_{ry\pm})$ have been introduced, with $\sigma$ being the Pauli operator.
The origin of this spin-orbit interaction
terms $\exp(i\sigma k_{rx}\delta^{H,V}_{rx\pm})$ and $\exp(i\sigma k_{ry}\delta^{H,V}_{ry\pm})$  lies in the transverse nature of the photon
polarization~\cite{Hosten2008}. The polarizations associated with
the plane-wave components experience different
rotations in order to satisfy the transversality
after reflection,
and hence induce geometric phases ( Berry phase )~\cite{Berry1984}.
The photonic Berry phase is closely related to the electronic Berry phase and spin-orbit coupling in graphene~\cite{Li2014II}.

\begin{figure}
\centerline{\includegraphics[width=8cm]{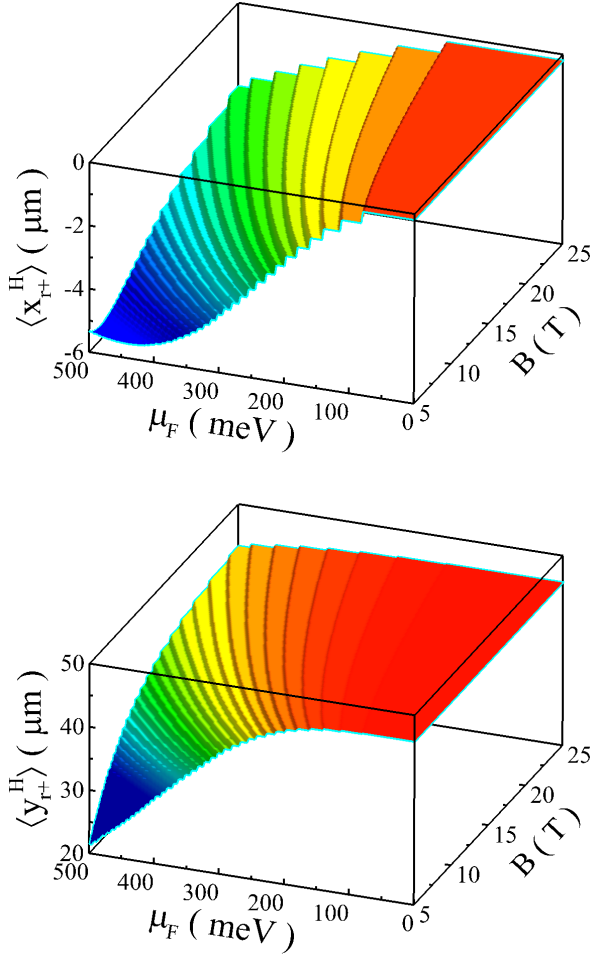}}
\caption{\label{Fig2} (Color online) Quantized in-plane (a) and transverse (b) spatial shifts
for graphene-substrate system as
a function of magnetic field and Fermi energy.
We assume the incident wavepacket with $|\mathrm{H}\rangle$ polarization, incident angle $\theta_i=60^\circ$, and $\omega/2\pi=1\mathrm{THz}$. The refractive index of undoped Si
in the terahertz range is $n_{Si}=3.415$.
We choose a carrier density of $4\times10^{16}\mathrm{cm}^{-3}$ and a mobility of
$1500\mathrm{cm^2/Vs}$. The temperature is chosen as $T=4\mathrm{K}$. These parameters are the same as in Ref.~\cite{Kamp2016}.}
\end{figure}

The spatial shifts of wave-packet at initial position ($z_r=0$) are given by
\begin{equation}
\langle{x_{r\pm}^{H,V}}\rangle=\frac{\langle\Phi_r^{H,V}|\partial_{k_{rx}}|\Phi_r^{H,V}\rangle}{\langle\Phi_r^{H,V}|\Phi_r^{H,V}\rangle}\label{PXHV},
\end{equation}
\begin{equation}
\langle{y_{r\pm}^{H,V}}\rangle=\frac{\langle\Phi_r^{H,V}|\partial_{k_{ry}}|\Phi_r^{H,V}\rangle}{\langle\Phi_r^{H,V}|\Phi_r^{H,V}\rangle}\label{PYHV}.
\end{equation}
Substituting Eqs.~(\ref{WPHII}) and (\ref{WPVII}) into Eqs.~(\ref{PXHV}) and (\ref{PYHV}), respectively,
the in-plane and transverse displacements of the two spin components can be written as
  \begin{equation}
  \langle x^{H}_{r\pm}\rangle= \mp\mathrm{Re}\left[\frac{\partial r_{sp}/\partial \theta_i}{k_0 (r_{pp}\mp ir_{sp})}\right]\label{XHRPM},
  \end{equation}
  \begin{equation}
  \langle y^{H}_{r\pm}\rangle= \mp\mathrm{Re}\left[\frac{(r_{pp}+r_{ss})\cot\theta_i}{k_0 (r_{pp}\mp ir_{sp})}\right]\label{YHRPM},
  \end{equation}
  \begin{equation}
  \langle x^{V}_{r\pm}\rangle=\mp\mathrm{Re}\left[\frac{-\partial r_{ps}/\partial \theta_i}{k_0 (r_{ss}\pm i r_{ps})}\right]\label{XVRPM},
  \end{equation}
  \begin{equation}
   \langle y^{V}_{r\pm}\rangle= \mp\mathrm{Re}\left[\frac{(r_{pp}+r_{ss})\cot\theta_i}{k_0 (r_{ss}\pm i r_{ps} )}\right]\label{YVRPM}.
  \end{equation}
By introducing the approximations $r_{pp}\gg{r_{sp}}$, $r_{pp}\gg{r_{ps}}$, $r_{pp}\simeq{R_p}$, $r_{ss}\simeq{R_s}$, $r_{ps}=r_{ps}=\sigma_{H}\sqrt{\mu_0/\varepsilon_0}(R_p-R_s)({\varepsilon/\varepsilon_0}-1)$, where
$R_p$ and $R_s$ are the Fresnel reflection coefficients at air-substrate interface, we find that the transverse shifts only depend on the optical properties of substrate, and therefore are not quantized.
These results coincide well with previous reported results~\cite{Kamp2016}.
In genernal, both the in-plane and transverse shifts in photonic SHE are quantized due to
the quantized Hall conductivity. In addition, the two spin components exhibit different absorption peaks in frequency space~\cite{Li2013II,Li2015},
and therefore an asymmetric spin-dependent splitting would occur in position space.

Figure~\ref{Fig2} shows the in-plane and transverse spatial shifts for the $|{H}\rangle$ polarization impinging on the graphene-substrate system. The quantized in-plane and transverse spatial shifts are plotted as a function of magnetic field $\mathbf{B}$ and Fermi energy $\mu_F$.
In a region of strong magnetic field, obvious quantitative steps are observed.
The corresponding plateaus become wider with the increase of the magnetic field $\mathbf{B}$ or the decrease of the Fermi energy $\mu_F$.
For the in-plane spatial shift, each saltus step almost has the same value
as shown in Fig~\ref{Fig2}(a). Comparing with the in-plane shifts, the transverse shifts exhibit lager values
as shown in Fig~\ref{Fig2}(b). For the same Fermi energy, the magnitude of the jump depends on the magnetic field.
When the magnetic field is strong enough, the spatial shifts and quantitative steps tend to zero.

We next consider the angular shifts in the photonic SHE. In the momentum-space representation, the angular shift can be written as
\begin{equation}
\langle{\Theta_{rx\pm}^{H,V}}\rangle=\frac{1}{k_0}\frac{\langle\Phi_r^{H,V}|{k_{rx}}|\Phi_r^{H,V}\rangle}{\langle\Phi_r^{H,V}|\Phi_r^{H,V}\rangle},\label{AXHV}
\end{equation}
\begin{equation}
\langle{\Theta_{ry\pm}^{H,V}}\rangle=\frac{1}{k_0}\frac{\langle\Phi_r^{H,V}|{k_{ry}}|\Phi_r^{H,V}\rangle}{\langle\Phi_r^{H,V}|\Phi_r^{H,V}\rangle}.\label{AYHV}
\end{equation}
Substituting Eqs.~(\ref{WPHII}) and (\ref{WPVII}) into Eqs.~(\ref{AXHV}) and (\ref{AYHV}), respectively,
we obtain the in-plane and transverse angular shifts for two spin components:
\begin{equation}
\langle \Theta_{rx\pm}^{H}\rangle= \mp\frac{1}{z_R}\mathrm{Im}\left[\frac{\partial r_{sp}/\partial \theta_i}{k_0 (r_{pp}\mp ir_{sp})}\right]\label{TXHRPM},
\end{equation}
\begin{equation}
\langle \Theta_{ry\pm}^{H}\rangle= \mp\frac{1}{z_R}\mathrm{Im}\left[\frac{(r_{pp}+r_{ss})\cot\theta_i}{k_0 (r_{pp}\mp ir_{sp})}\right]\label{TYHRPM},
\end{equation}
\begin{equation}
\langle \Theta_{rx\pm}^{V}\rangle= \mp\frac{1}{z_R}\mathrm{Im}\left[\frac{-\partial r_{ps}/\partial \theta_i}{k_0 (r_{ss}\pm i r_{ps})}\right]\label{TXVRPM},
\end{equation}
\begin{equation}
\langle \Theta_{ry\pm}^{V}\rangle= \mp\frac{1}{z_R}\mathrm{Im}\left[\frac{(r_{pp}+r_{ss})\cot\theta_i}{k_0 (r_{ss}\pm i r_{ps} )}\right]\label{TYVRPM},
\end{equation}
where $z_R=k_0w_0^2/2$ is the Rayleigh length. Similar to the spatial shifts, both the in-plane and transverse angular shifts are quantized.
Significantly different from the spatial shifts, the angular shifts are spin independent.

\begin{figure}
\centerline{\includegraphics[width=8cm]{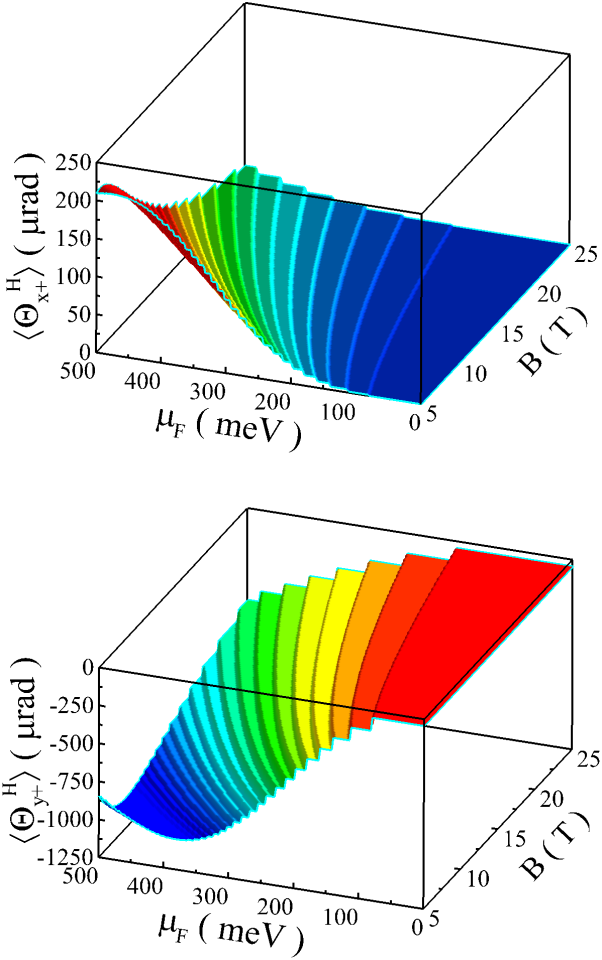}}
\caption{\label{Fig3} (Color online) Quantized in-plane (a) and transverse (b) angular shifts in photonic SHE as a function of magnetic field and Fermi energy. The beam waist is chosen as $w_0=1\mathrm{mm}$.
Other parameters are the same as in Fig.~\ref{Fig2}.}
\end{figure}

Figure~\ref{Fig3} shows the in-plane and transverse angular shifts.
In the quantum Hall regime, the quantized effect of angular shifts
can be clearly observed and the plateaus become wider with the increase of the magnetic field $\mathbf{B}$
or the decrease of the Fermi energy $\mu_F$.
Comparing with Figs.~\ref{Fig3}(a) and \ref{Fig3}(b) shows that the transverse angular shifts exhibit larger quantitative steps.
Similar to the situation in spatial shifts, the magnitude of the jump of angular shifts depends on the magnetic field $\mathbf{B}$.
When the magnetic field is strong enough, the angular shifts and the quantitative steps tend to zero.

We now examine the role that Hall conductivity plays in the quantized photonic SHE.
Substituting Eqs.~(\ref{RPS}) and (\ref{HallC}) into Eqs.~(\ref{XHRPM}) and (\ref{TXHRPM}), the in-plane spatial and angular shifts can be written as quantized functions of the magnetic field
\begin{equation}
 \langle x^{H,V}_{r\pm}\rangle=\pm2(2n_c+1)\mathrm{Sgn}[B]\frac{e^2}{4\pi\hbar{\varepsilon_0}c}\frac{1}{k_0}\mathrm{Re}[W^{H,V}]\label{PXHVIII},
\end{equation}
\begin{equation}
\langle{\Theta_{rx\pm}^{H,V}}\rangle=\pm2(2n_c+1)\mathrm{Sgn}[B]\frac{e^2}{4\pi\hbar{\varepsilon_0}c}\frac{1}{k_0z_R}\mathrm{Im}[W^{H,V}]\label{AXHVIII},
\end{equation}
Here, ${e^2}/{4\pi\hbar{\varepsilon_0}c}$ is the fine structure constant and $W^{H, V}$ can be written as
\begin{equation}
W^{H}=\frac{4}{r_{pp}\mp ir_{sp}}\frac{\partial[(k_{iz}k_{tz})/(\alpha^T_+\alpha_+^L+\beta)]}{\partial\theta_i},\label{WH}
\end{equation}
\begin{equation}
W^{V}=\frac{4}{r_{ss}\mp ir_{ps}}\frac{\partial[(k_{iz}k_{tz})/(\alpha^T_+\alpha_+^L+\beta)]}{\partial\theta_i}.\label{WV}
\end{equation}
Equations~(\ref{PXHVIII}) and (\ref{AXHVIII}) show that the in-plane spatial and angular shifts in photonic SHE
are quantized in integer multiples of the fine structure constant under a certain condition: $r_{pp}\gg{r_{sp}}$ and $r_{ss}\gg{r_{ps}}$ for $|{H}\rangle$ polarization and $|{V}\rangle$ polarization, respectively.
These quantized behaviors are attributed to the quantized Hall conductivity.

\section{Quantum weak measuments}
We propose the signal enhancement
technique known as quantum weak
measurements~\cite{Aharonov1988,Ritchie1991,Dressel2014} to detect the quantized photonic SHE.
After the pre-selection of state, weak interaction, and post-selection of state, the
wavefunction evolves to the final state
\begin{eqnarray}
 |{\Phi}_{f}\rangle
 &=& \langle \psi_f|\exp(i\sigma k_{rx}\delta_{rx})\exp(i\sigma k_{ry}\delta_{ry})|\psi_i\rangle |{\Phi}\rangle\nonumber\\
 &=& \langle \psi_f|1+i\sigma k_{rx}\delta_{rx}+i\sigma k_{ry}\delta_{ry}|\psi_i\rangle |{\Phi}\rangle\nonumber\\
 &\approx&\langle \psi_f|\psi_i\rangle \left[1+(ik_{rx}\delta_{rx}+ik_{ry}\delta_{ry}) \frac{\langle \psi_f|\sigma|\psi_i\rangle}{\langle \psi_f|\psi_i\rangle} \right]|{\Phi}\rangle\nonumber\\
 &=& \langle \psi_f|\psi_i\rangle(1+ik_{rx}A_w\delta_{rx}+ik_{ry}A_w\delta_{ry})|{\Phi}\rangle.
\end{eqnarray}
Here, $A_w$ is so called weak value which can be written as
\begin{equation}
A_{w}=\frac{\langle \psi_f|\sigma|\psi_i\rangle}{\langle
\psi_f|\psi_i\rangle},\label{AWI}
\end{equation}
where $\sigma$ is the Pauli operator, $|\psi_i\rangle$ and $|\psi_f\rangle$ are the
polarization states of the pre-selection and post-selection, respectively.
In a general case, both $A_w$ and
$\delta^{H,V}_{rx}$ are complex. However, only the imaginary parts of $A_w\delta^{H,V}_{rx}$ and $A_w\delta^{H,V}_{ry}$  can
be amplified and are given by
\begin{equation}
\mathrm{Im}[A_w \delta_{rx}^{H,V}]
=\mathrm{Re}[A_w]\mathrm{Im}[\delta_{rx}^{H,V}]+\mathrm{Im}[A_w]\mathrm{Re}[\delta_{rx}^{H,V}]\label{AWDX},
\end{equation}
\begin{equation}
\mathrm{Im}[A_w \delta_{ry}^{H,V}]
=\mathrm{Re}[A_w]\mathrm{Im}[\delta_{ry}^{H,V}]+\mathrm{Im}[A_w]\mathrm{Re}[\delta_{ry}^{H,V}]\label{AWDY}.
\end{equation}
Here, the imaginary and real parts of shifts are determined
by the real and imaginary parts of the weak values, respectively.

\begin{figure}
\centerline{\includegraphics[width=8cm]{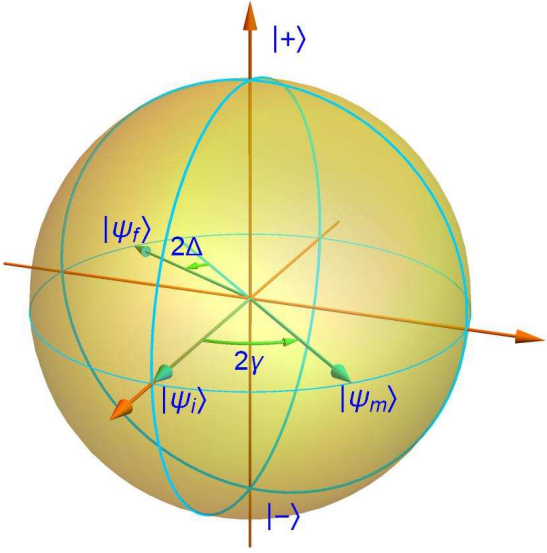}}
\caption{\label{Fig4} (Color online) Representation of the pre-selection
and post-selection states for $|\psi_i\rangle$ and $|\psi_f\rangle$ on the Poincar\'{e} sphere. A middle state $|\psi_m\rangle$
is also introduced due to the Kerr rotation. The quantum state is chosen as $\Theta=\pi/2$.  For $|{H}\rangle$ input
polarization $\Phi=0$, while for $|{V}\rangle$
input polarization $\Phi=\pi$. The angle $\Delta$ give origin to the real and imaginary parts of the weak value.}
\end{figure}

To obtain a clear picture, we represent the pre-selection and post-selection states on Poincar\'{e} sphere as shown in Fig.~\ref{Fig4}.
The preselection state can be written as
\begin{equation}
|\psi_i\rangle=\cos\left(\frac{\Theta}{2}\right)|+\rangle+e^{-i\Phi}\sin\left(\frac{\Theta}{2}\right)|-\rangle,\label{PRES}
\end{equation}
where $0\leq\Theta\leq\pi$ and $0\leq\Phi\leq{2\pi}$ represent quantum state on the
Poincar\'{e} sphere~\cite{Jordan2014}. For $|{H}\rangle$ input
polarization $\Theta=\pi/2$ and $\Phi=0$, while for $|{V}\rangle$
input polarization $\Theta=\pi/2$ and $\Phi=\pi$.
Due to the Kerr effect, we also introduce a middle polarization state:
\begin{equation}
|\psi_m\rangle=\cos\left(\frac{\Theta}{2}\right)|+\rangle+e^{-i(\Phi+2\gamma)}\sin\left(\frac{\Theta}{2}\right)|-\rangle\label{MIDS}.
\end{equation}
Here, $\gamma=\arctan(r_{ps}/r_{pp})$  and $\gamma=\arctan(r_{sp}/r_{ss})$ are the Kerr rotation angles for $|{H}\rangle$ and $|{V}\rangle$ polarization, respectively. After the post-selection, the final polarization state can be written as
\begin{equation}
|\psi_f\rangle=\cos\left(\frac{\Theta}{2}\right)|+\rangle+e^{-i\left(\Phi+2\gamma+\pi+2\Delta\right)}\sin\left(\frac{\Theta}{2}\right)|-\rangle,\label{POSTS}
\end{equation}
where $\Delta$ is the postselection angle.
In the commonly used weak measurements system, the two
polarization states $|\psi_i\rangle$ and $|\psi_f\rangle$ are nearly perpendicular to each other~\cite{Aiello2008,Qin2009,Luo2011II}.
In our case, however, the condition no longer holds due to the Kerr rotation of polarization state.
Therefore, a middle polarization state is introduced and the weak value $A_w$ should be modified as
\begin{equation}
A_{w}=\frac{\langle \psi_f|\sigma|\psi_m\rangle}{\langle
\psi_f|\psi_m\rangle}.\label{AWI}
\end{equation}
Substituting Eqs.
(\ref{MIDS}) and (\ref{POSTS}) into Eq.~(\ref{AWI}), the
weak value can be obtained as
\begin{equation}
\mathrm{Re}[A_{w}]=0,~~~\mathrm{Im}[A_{w}]=\cot\Delta.\label{AWII}
\end{equation}
Here, the quantum state with $\Theta=\pi/2$ has been introduced.

We now consider the amplified shift in the farfield. After
a free evolution, the wave-packet moves to its final position
\begin{equation}
\langle{x^{H,V}_w}\rangle=\frac{z_r}{k_0}\frac{\langle\Phi_f|k_{rx}|\Phi_f\rangle}{\langle\Phi_f|\Phi_f\rangle}\label{AXHVII},
\end{equation}
\begin{equation}
\langle{y^{H,V}_w}\rangle=\frac{z_r}{k_0}\frac{\langle\Phi_f|k_{ry}|\Phi_f\rangle}{\langle\Phi_f|\Phi_f\rangle}\label{AYHVII}.
\end{equation}
When $\delta_{rx}^{H,V}$ and $\delta_{ry}^{H,V}$ are complex, it is difficult to
distinguish and detect the spatial shift and angular shift, since they have
combined contribution to the amplified shift in far field.

Here, we just consider the case that the weak value is purely imaginary. From Eqs.~(\ref{AXHVII}) and (\ref{AYHVII}) the
amplified shifts for $|{H}\rangle$ and $|{V}\rangle$ polarization
are given by
  \begin{equation}
 \langle{x^{H,V}_{w}}\rangle=\frac{z_{r}}{z_R}\mathrm{Re}[\delta^{H,V}_{rx}]\cot\Delta
 =\frac{z_{r}}{z_R}|\langle x^{H,V}_{r\pm}\rangle|\cot\Delta\label{ASSS},
 \end{equation}
 \begin{equation}
 \langle{y^{H,V}_{w}}\rangle=\frac{z_{r}}{z_R}\mathrm{Re}[\delta^{H,V}_{ry}]\cot\Delta
 =\frac{z_{r}}{z_R}|\langle y^{H,V}_{r\pm}\rangle|\cot\Delta\label{ASSS}.
 \end{equation}
Here, the weak value only amplifies the spatial shift. For $z_r=1000\mathrm{mm}$ and $\Delta=1^\circ$, the spatial shifts at the detector can be obtained by an amount that is about
$5470$ times larger than the initial spatial shift caused by the photonic SHE. For in-plane and transverse spatial shifts,
the quantized jumps are $361\mathrm{nm}$ and $96\mathrm{nm}$, respectively
($B=12\mathrm{T}$ and $\mu_F=200\mathrm{meV}$).
The corresponding amplified quantized jumps can be obtained as $1.97\mathrm{mm}$ and $0.53\mathrm{mm}$ in the quantum weak measurements.
A large amplifying factor can be obtained as $\Delta\rightarrow0$.
Under this condition, a modified weak measurement should be developed~\cite{Zhou2014,Chen2015}.
Based on the recent advances in terahertz technology, such as terahertz lasers, detectors, and terahertz elements~\cite{Tonouchi2007}, we believe the experimental observation of the quantized photonic SHE is possible.

\section{Conclusions}
In conclusion, we have revealed a quantized photonic SHE in quantum Hall regime of graphene.
The quantized photonic SHE manifests itself as quantized in-plane and transverse spin-dependent splittings due to the quantized Berry phase
which is related to the quantized spin-orbit interaction.
We have demonstrated that the in-plane spin-dependent splitting is quantized in
multiples of the fine structure constant,
which is attributed to the quantized Hall conductivity of graphene.
Furthermore, a signal enhancement
technique known as quantum weak measurement has been theoretically
proposed to detect the quantized photonic SHE.
It should be noted that the quantized photonic SHE is different from the quantum SHE of light
--surface modes with
strong spin-momentum locking~\cite{Bliokh2014,Bliokh2015II,Mechelen2016}.
By incorporating the quantum weak measurement techniques, the quantized photonic SHE
holds great promise for detecting quantized Hall conductivity and Berry phase.
Therefore, we believe that these results may bridge the gap
between the electronic SHE and photonic SHE in graphene.

\begin{acknowledgements}
This research was supported by the National Natural Science Foundation
of China (Grants Nos. 11274106 and 11474089).
\end{acknowledgements}

\end{document}